\documentclass[twoside]{article}
\usepackage{graphicx} 
\usepackage{subfigure}
\usepackage{xspace}
\usepackage{url}
\usepackage{algorithm}
\usepackage{algorithmic}
\usepackage{amsfonts}
\usepackage{amsmath}
\usepackage{fancyhdr}
\usepackage{times}

\begin{document}

\newcommand{\ieee}{\textsc{ieee}~802.11\xspace}
\newcommand{\brackets}[1]{$\angle(#1)$}
\newcommand{\sector}[1]{$\mathbb{S}\left(#1\right)$}
\newcommand{\proxgraph}{proxi-graph\xspace}

\newenvironment{fig}{\begin{figure}[h!tbp]}{\end{figure}}

\title{Efficient Greedy Geographical Non-Planar Routing with Rreactive Deflection}
\author{
Fabrice Theoleyre, Eryk Schiller and Andrzej Duda
\vspace{0.4cm}\\
{\small Grenoble Informatics Laboratory}
\\
{\small CNRS and Grenoble-INP}
\\
{\small 681 rue de la passerelle, BP72}
\\
{\small 38402 Saint Martin d'Heres, France}
\\
{\small Email: \url{\{firstname.lastname\}@imag.fr}}
}

\maketitle

\begin{abstract}
We present a novel geographical routing scheme for spontaneous wireless mesh networks. Greedy geographical routing has many advantages, but suffers from packet losses occurring at the border of \emph{voids}. In this paper, we propose a flexible greedy routing scheme that can be adapted to any variant of geographical routing and works for any connectivity graph, not necessarily Unit Disk Graphs. The idea is to reactively detect voids, backtrack packets, and propagate information on blocked sectors to reduce packet loss. We also propose an extrapolating algorithm to reduce the latency of void discovery and to limit route stretch. Performance evaluation via simulation shows that our modified greedy routing avoids most of packet losses.
\end{abstract} 



\section{Introduction}

We consider wireless mesh networks composed of a large number of
wireless routers providing connectivity to mobile nodes. They begin to emerge
in some regions to provide cheap network connectivity to a community
of end users. Usually they grow in a {\em spontaneous} way when
users or operators add more routers to increase capacity and coverage.

We assume that mesh routers benefit from abundant resources (memory,
energy, computation power, GPS devices in some cases), may only move,
quit, or join occasionally, so that the topology of a typical mesh
networks stays fairly stable.  The organization of mesh networks needs
to be {\em autonomic}, because unlike the current Internet, they
cannot rely on highly skilled personnel for configuring, connecting,
and running mesh routers. Spontaneous growth of such networks may
result in a dense and unplanned topology with some uncovered areas.

Unlike traditional approaches, {\em geographical routing} presents
interesting properties for spontaneous wireless mesh networks: it does
not require any information on the global topology since a node
choses the next hop among its neighbor routers based of the
destination location. Consequently, the routing scheme is scalable,
because it only involves local decisions. Geographical
routing is simple, because it does not require routing tables so that
there is no overhead of their creation and maintenance.  Joining the
network is also simple, because a new mesh router only needs an
address based on its geographical position.  Such addresses can be
obtained from a dedicated device (e.g. GPS) or with methods for
deriving consistent location addresses based on the information from
neighboring nodes about radio signal strength \cite{capkun02} or
connectivity \cite{rao03}.
The most familiar variant of geographical routing is {\em greedy
  forwarding} in which a node forwards a packet to the neighbor
closest to the destination \cite{bose99,karp00b}. Greedy forwarding
guarantees loop-free operation, but packets may be dropped at {\em
  blocked nodes} that have only neighbors in the backward
direction. Blocked nodes appear at some places near uncovered areas
(\emph{voids}) or close to obstacles to radio waves in a given direction.

Our main contribution is to propose a new greedy routing that
correctly deals with voids. First, we define a new
mechanism to reactively detect voids and surround them, which
significantly reduces packet loss. Moreover, the information of
detected voids propagates backwards so that subsequent packets to the
same direction benefit from this reactive detection. Second, we
propose a mechanism in which voids deviate packets and shorten the
length of a route compared to classical approaches. Our routing scheme
works in any network topology independently of whether it corresponds
to a planar graph or not.

We start with the description of the related work on geographical
routing in Section~\ref{section:relatedwork}.
Section~\ref{section:new_greedy_routing} presents the details of the
proposed new greedy routing protocol. Then, we evaluate its
performance via simulation in Section~\ref{section:results} and
conclude.


\section{Related Work}
\label{section:relatedwork}

Geographic information can largely reduce complexity of routing in
spontaneous mesh networks. The most simple and widely used protocol
is greedy geographic routing \cite{bose99,carlsson07,he06,casari07}: when a
node receives a packet, it uses the following forwarding rule:
\begin{quote}
  "forward the packet to the node with the best improvement".
\end{quote}
\emph{Improvement} is usually defined with respect to the distance
towards the destination. Since improvement is not negative, there is
no routing loops. Moreover, routing is scalable, because all routing
decisions are local.

Geographical routing requires addresses based on geographical
coordinates: a node must obtain its location either with a dedicated
physical device (e.g. GPS) or through a more complex algorithm,
e.g. by estimating the position with respect to its neighbors. Capkun
et al. propose to construct a local coordinate system for each node
and determine the coordinates of its neighbors \cite{capkun02}. Then,
they aggregate the local coordinate systems into global
coordinates. The authors assume the distance to each neighbor known,
but usually it is difficult to obtain. Niculescu et al. follow a
similar approach, but based on the \emph{angle of arrival} of packets
coming from neighbors \cite{niculescu03}.
A pragmatic approach to this problem is to assume that a subset of
mesh routers know their exact positions via GPS devices and other
nodes can compute their positions with respect to its neighbors
\cite{bulusu01}.

The main drawback of greedy geographical routing is packet loss at
blocked nodes near voids or obstacles. A node must drop a packet when
the improvement associated with any of its neighbors is negative
(cf. Figure~\ref{fig:void}). In \emph{face routing} the
\emph{left-hand} rule \cite{karp00b} tries to go around a void, but it
requires the connectivity graph of nodes to be planar. Relative
Neighborhood Graphs can yield planar graphs for Unit Disk Graphs
\cite{cartigny05}, but in real wireless environments, the conditions
for obtaining planar graphs are not satisfied due to asymmetric links
and not circular radio coverage \cite{kim06}. To the best of our
knowledge, there is no efficient and localized planarization algorithm
proposed for a general connectivity graph. A possible solution to this
problem is the following method: a border node initiates local
flooding to find the next hop closer to the destination
\cite{stojmenovic01}. However, it results in long delays and
significant overhead. Fotopoulou et al. \cite{fotopoulou04} propose to
adapt this method to establish and maintain a \emph{virtual
  circuit}. Funke et al. \cite{funke05} propose an algorithm inspired
by topological geometry to discover the void limits, i.e. the
\emph{border nodes} by creating isosets. However, this method does not
work for non Unit Disk Graphs since isosets are not rings in general
connectivity graphs. Very recently, \cite{arad09} proposed to maintain both virtual and physical coordinates. When a node detects that it is blocked, it changes its virtual position increasing its height to dissuade neighbors to send packets. Greedy routing will use virtual coordinates in order to surround the voids. However, a node is considered blocked only if an empty sector of more than 220¡ exists, and not all voids can be detected in this manner. 

We propose here a generic method to deal with voids by backtracking
packets and discovering blocked areas near voids: deflection routing
deviate packets outside them. Thus, the algorithm presented here is
perfectly supplementary to existing methods: deflection improves
greedy routing by trying to surround voids and any other technique
presented above can be used when a void is reached.

\begin{fig}
\begin{center}
	\includegraphics[width=9cm]{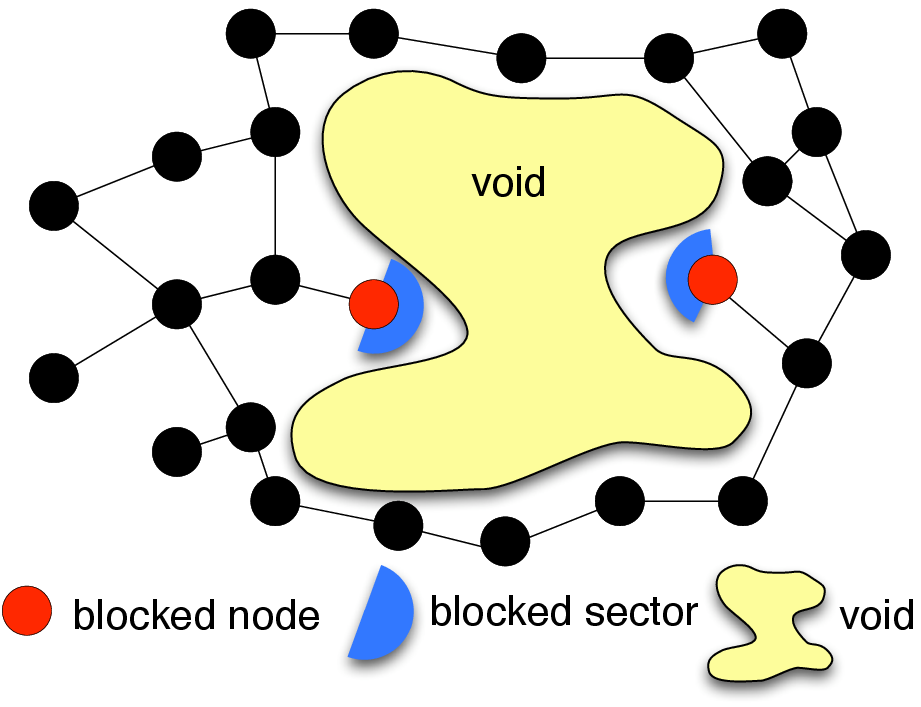}
	\vspace{-0.2cm}
	\caption{Blocked nodes in greedy geographical routing.}
	\label{fig:void}
\end{center}
\end{fig}


\section{Reactive Deflection}
\label{section:new_greedy_routing}

Geographical routing is attractive for mesh networks, but suffers from
two main drawbacks: blocked nodes can drop many packets and the route
length may drastically increase when a surrounding mechanism tries to
deviate a packet around a void (e.g. the left-hand rule in unit disk
graphs). In this paper, we assume a general connectivity graph and
propose to reactively detect blocked nodes and locally advertise
blocked sectors to avoid packet losses. Such a technique is efficient
in any type of networks and graphs since it does not assume any
particular graph property.

Detection of blocked nodes can be done in a proactive way: locally
flood information to detect voids. For example, we can discover the
topology of the wireless mesh to detect elementary cycles in which no
other node is located inside the ring. The location of nodes helps to
surround voids. However, such an approach requires a complete
knowledge of the mesh topology and is computationally intensive.

In opposition to this approach, we have chosen a reactive method: a
node becomes \emph{blocked} with respect to a given destination when
it cannot forward a packet to any neighbor closer to the
destination. Hence, the part of the network not concerned by
forwarding this packet does not generate any control traffic so that
this approach is more scalable.


Let us first adopt the following notation:
\begin{itemize}
	\item $d(A,B)$: the Euclidean distance between the geographical coordinates of nodes $A$ and $B$
  	\item \brackets{AB,AC}: the oriented angle between two coordinates of nodes $(A,B)$ and $(A,C)$ (by convention, we denote by \brackets{AB} the normalized angle \brackets{
\begin{pmatrix}
1\\
0
\end{pmatrix}
,AB} )
\item \sector{S,\alpha,\beta,d_{min}}: the sector of node $S$ composed
  of all nodes $N$ such that $\alpha \leq$\brackets{SN}$\leq \beta$
  and such that $d(S,N) \geq d_{min}$
\end{itemize}


In our approach, a node chooses a neighbor closer to the destination
and not blocked for this direction. If a node fails to forward a
packet to a given destination, it will consider itself as blocked for
this direction. It will advertise backwards a list of blocked
directions so that its neighbors will not choose it as a next hop for
these directions.  If several non blocked neighbors exist, the
forwarder chooses the neighbor closest to the destination, i.e. with
the best \emph{improvement}.

\begin{fig}
\begin{center}
	\includegraphics[width=0.8\linewidth]{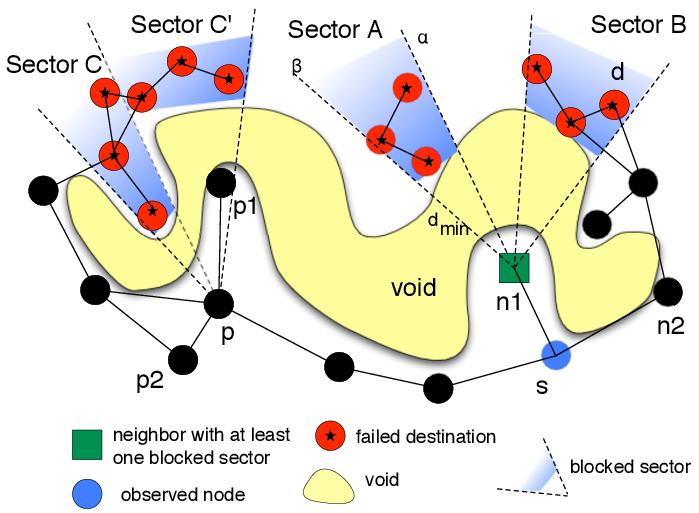}
	\vspace{-0.2cm}
	\caption{Examples of blocked nodes and blocked sectors.}
	\label{fig:blocked_sectors}
\end{center}
\end{fig}

For advertising blocked directions, we propose to use the notion of
\emph{blocked sectors}: a node $N$ advertises that it is blocked for
any destination that falls in sector
\sector{N,angle_{min},angle_{max},dist_{min}}. Let us consider the
topology illustrated in Figure~\ref{fig:blocked_sectors}. Node $n_1$
wants to forward a packet to destination $d$ and it discovers that it
is blocked for this destination since no neighbor exists in this
direction. Thus, it backwards the packet to $s$ with its two blocked
sectors. Based on this information, $s$ marks $n_1$ as blocked and
forwards the packet to another neighbor closer to $d$ (node $n_2$ in
this case). 

\begin{algorithm}
	\begin{algorithmic}[1]
		\STATE $next \gets \emptyset$
		
		\FORALL{$n \in Neighbors$}
			\IF{$d(n,D) < d(N,D)$ \textbf{and} !{\sc Blocked}(n,D) \textbf{and} $d(n,D) < d(next,D)$}
				\STATE $next \gets n$
			\ENDIF
		\ENDFOR
		\IF{$next = \emptyset$}
		\STATE
			Blocked(N,D) $\gets$ \textbf{true}\\
			$next \gets previous\_hop$
		\ENDIF
		\STATE \textbf{return} $next $
	\end{algorithmic}
\caption{ReactiveDeflection(N$\rightarrow$D)}
\label{algo:greedy_routing}
\end{algorithm}

To limit the overhead, a node tries to merge all its blocked sectors
before advertising them. It can only merge overlapping sectors having
the same minimal distances (within some tolerance $\Delta_d$).
Otherwise, the merged blocked sector may include nodes that are
reachable---consider for instance the topology of figure~\ref{fig:blocked_sectors}: if node $p$  merges sectors $C$ and $C'$, node $p_1$ may appear in the blocked sector. Thus, it would become unreachable from $p_2$. Clearly, we must avoid such a merging. Only sectors will the same $d_{min}$ will be merged : tolerance $\Delta_d$ allows some merging of sectors with approximately equal minimal distances.


More formally, node $N$ executes Algorithm \ref{algo:greedy_routing}. Procedure $ReactiveDeflection()$ finds the next hop for forwarding a packet to destination $D$: the next hop must be closer to the destination and must be unblocked for $D$. If it does not return any node, it means that node $N$ becomes \emph{blocked} for destination $D$ (variable {\sc Blocked}(N,D) becomes true).  Thus, node $N$ updates its blocked sectors and sends the packet backwards to the previous hop with its list of blocked sectors piggybacked onto the packet. This scheme is loop-free: when a node sends a packet backwards, the receiver will update its blocked sectors and it cannot choose the same next hop for subsequent packets, because Algorithm~\ref{algo:greedy_routing} does not forward packets to blocked nodes.

\begin{fig}
\begin{center}
	\includegraphics[width=0.8\linewidth]{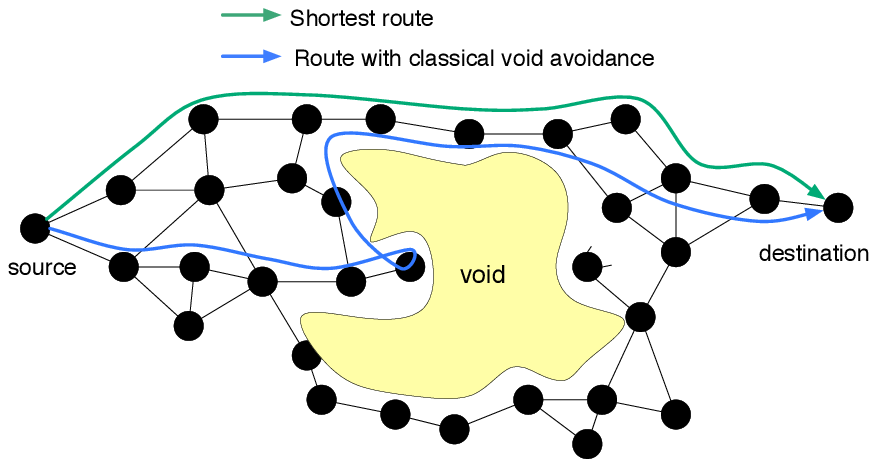}
	\vspace{-0.2cm}
	\caption{The increase of route length when surrounding a void.}
	\label{fig:voids_route_length}
\end{center}
\end{fig}


In networks with non unit disk graph topology, when a node becomes blocked and there are no other neighbor closer to the destination, the node needs to discover a node in a larger vicinity able to forward the packet to the destination. Usually, it consists of flooding a request in a $k$-neighborhood of the node, $k$ being a parameter to limit the scope of flooding.  In this case, the length of the route may increase, which is illustrated in Figure~\ref{fig:voids_route_length}:
border nodes need to forward the packet to reach a \emph{virtual} next hop \cite{bose99,fotopoulou04}. This increases both the load of the border nodes and the route length. We propose to limit the effect of such a behavior.

Note that when we reduce packet loss with the previously described algorithm, we also reduce in the long term the route length. Indeed, the nodes around the void discover that they have blocked sectors. When they propagate the information about blocked sectors, nodes with all blocked neighbors also become blocked for this destination. Finally, each node discovers a blocked area and forwards packets outside this area. However, we need several useless packet transmissions and backtracking before the network converges, and blocked sectors are correctly constructed. We propose a mechanism to accelerate the convergence of this propagation process by extrapolating the location of a blocked area.


\begin{fig}
\begin{center}
	\includegraphics[width=0.7\linewidth]{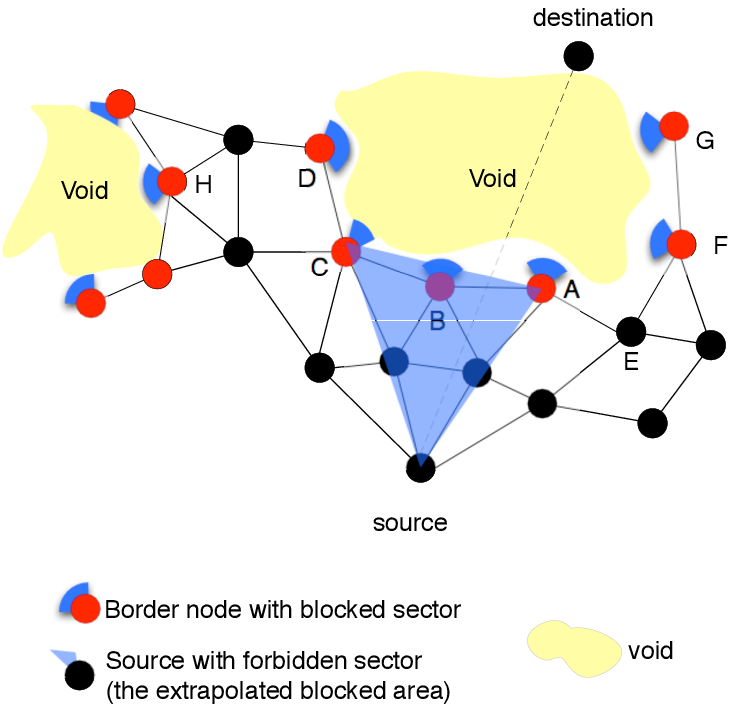}
	\vspace{-0.2cm}
	\caption{Method for locally detecting the border of a void.}
	\label{fig:voids_limits}
\end{center}
\end{fig}

We propose to detect the border of a void based on only local
neighborhood knowledge. We will show that even if a node has only
local knowledge, i.e. about nodes at a limited distance, voids can be
efficiently surrounded. When a node must transmit a packet backwards,
it locally floods an \texttt{hello} packet containing the list of its
neighbors and blocked sectors in a $k$ hop scope. 

To detect the border of a void, node $N$ first searches for the
blocked k-neighbor closest to the direction of the destination $D$,
i.e. minimizing angle \brackets{(N,D), (N,BN)} for all blocked nodes
$BN$. Then, $N$ constructs the Maximum Connected Set of blocked nodes
that contains $BN$: it adds $BN$ to this set, and recursively adds all
its blocked neighbors. Finally, $N$ computes the \emph{forbidden
  sector} that spans the maximum connected set---it extrapolates the
blocked area.

Figure~\ref{fig:voids_limits} illustrates void detection with
the knowledge of the 3-neighborhood topology. First, the
source node detects if it knows a node with a blocked direction and takes
the closest one to the direction to the destination. In the example the blocked node is $B$. Then, the source constructs the connected
set of blocked nodes that includes node $B$: it obtains set
$\{A,B,C,D\}$. Obviously, node $F$ is not present in the set since
it is not connected to $A$ via other blocked nodes. In the same way,
border node $H$ is not in  set $\{A,B,C,D\}$, because it is 2
hops away: it is border to another void. Finally, we obtain the
forbidden sector for the destination. We can note that
node $E$ is not blocked since it can choose node $F$ when $A$ is
blocked: $E$ will never be blocked for the direction.


Algorithm~\ref{algo:greedy_routing_blocked_sectors} presents the modified protocol.  Function {\sc IsInForbiddenSector} computes the forbidden sector and returns {\sc true}, if the node is located inside this sector. Function {\sc CloserToSectorLimits(P,Q)} returns {\sc true}, if $P$ is closer than $Q$ to the forbidden sector limits.

\begin{algorithm}
	\begin{algorithmic}[1]
		\STATE $next \gets \emptyset$
		
		\FORALL{$n \in Neighbors$}
			\IF{$d(n,D) < d(S,D)$  \textbf{and} !{\sc Blocked}(n,D)}
				\IF{!{\sc isInForbiddenSector}(n)  \textbf{and} \{ {\sc isInForbiddenSector}(next) \textbf{or} $d(n,D) < d(next,D)$ \} }
				\STATE $next \gets n$
				\ELSIF{{\sc isInForbiddenSector}(next) \textbf{and} {\sc IsInForbiddenSector}(n) \textbf{and} {\sc CloserToSectorLimits}(n,next)}
				\STATE $next \gets n$
				\ENDIF
			\ENDIF
		\ENDFOR
		\IF{$next = \emptyset$}
		\STATE
			Blocked(N,D) $\gets$ \textbf{true}\\
			$next \gets previous\_hop$
		\ENDIF
		\STATE \textbf{return} $next $
	\end{algorithmic}
\caption{ModifiedReactiveDeflection(D,ForbiddenSector)}
\label{algo:greedy_routing_blocked_sectors}
\end{algorithm} 

In other words, if some next hops exist and do not lie in the computed forbidden sector, we choose the best one. Otherwise, if all possible next hops are in this forbidden sector, we choose the node closest to the limits of the forbidden sector.  With this modified routing scheme, we forward packets outside the forbidden sector, because a void appears as something repellent to packets by creating forbidden sectors in a distributed manner while keeping routing loop-free.


\section{Performance evaluation}
\label{section:results}

\begin{table}
\begin{center}
\caption{Route length with deflection routing for different values of the k-neighborhood}
\label{tab:kvalues}
\begin{tabular}{|c|c|}
	\hline
	$k$ (in hops)		& Route length \\
	\hline
	1	& 	33.6\\
	2	& 	33\\
	3	& 	32.7\\
	4	& 	33.2\\
	5	& 	33.6\\
	\hline
\end{tabular}
\end{center}
\end{table}
We have generated random meshes of 1000 nodes according to two models: Unit Disk Graphs \cite{clark90} and what we call a \emph{\proxgraph} (a graph based on proximity). In a \proxgraph, each node chooses a \emph{radio range} following a Gaussian distribution centered at 1 with standard deviation $Std$ depending on the radio range (we assume
$Std=25\% \cdot (radio-range)$ in our simulations). We consider a \proxgraph with a rectangular void of size two fifth of the simulation disk radius in the center of the simulation area.  Besides, we discard disconnected topologies and use a disk simulation area to reduce border effects. Data traffic consists of 1,000 flows of 10 packets each from a random source to a random destination. 

At the beginning, to evaluate only the properties of routing itself, we assume ideal radio and MAC layers: packets do not experience any loss due to channel or MAC behavior to only test routing properties. Then, we evaluate the performances of the proposed protocols with the ns2 simulator to take into account more realistic radio conditions.
Finally, we assume that nodes advertise the list of blocked sectors to their neighbors and a node is aware of the blocked nodes in its 3-neighborhood (hellos contain the list of neighbors) since it achieves the best tradeoff between the performance and the overhead as shown in the simulations.

\begin{fig}
\begin{center}
		\includegraphics[width=9cm]{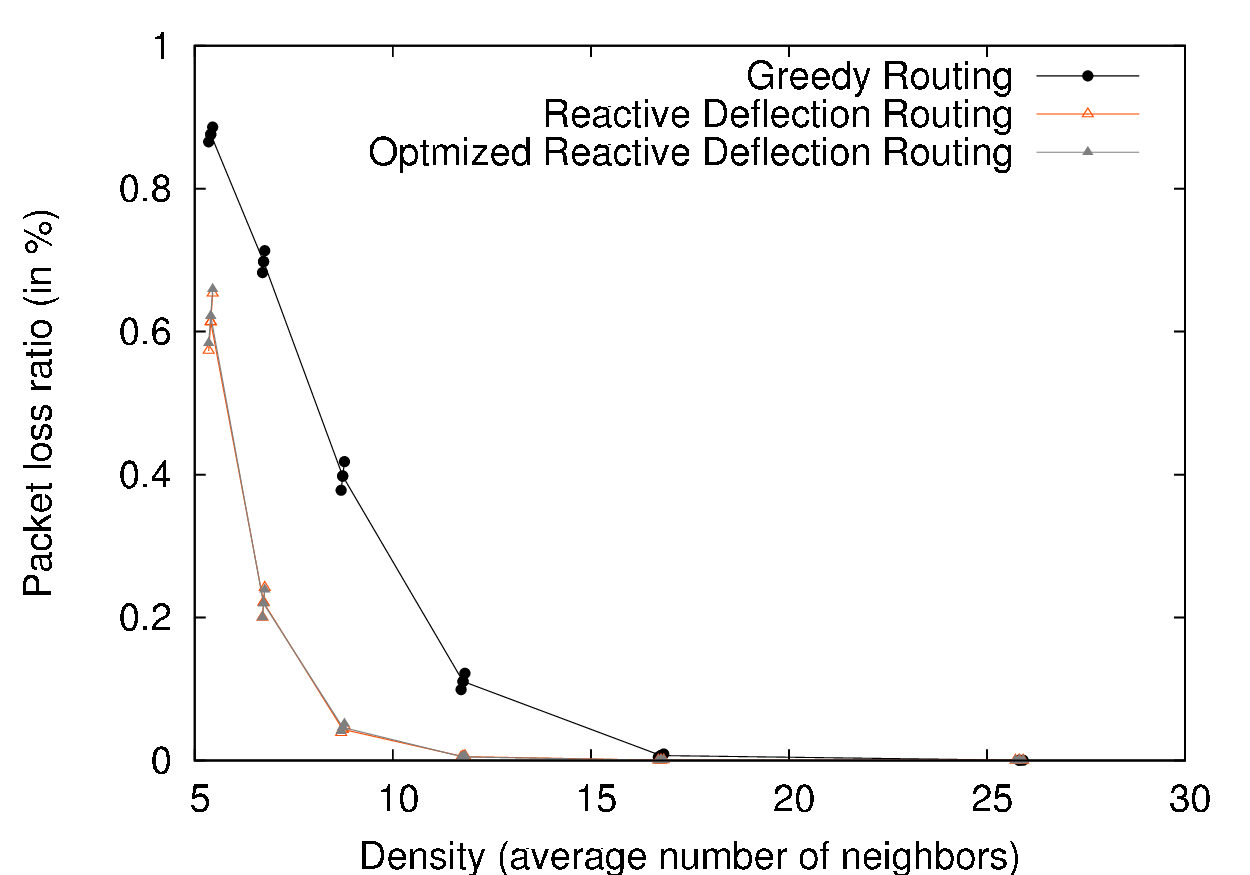}
	\vspace{-0.2cm}
	\caption{Packet loss under different routing schemes for UDG.}
	\label{fig:udg-packet_loss}
\end{center}
\end{fig}

\begin{fig}
\begin{center}
	\subfigure[Route length in hops]{
		\includegraphics[width=9cm]{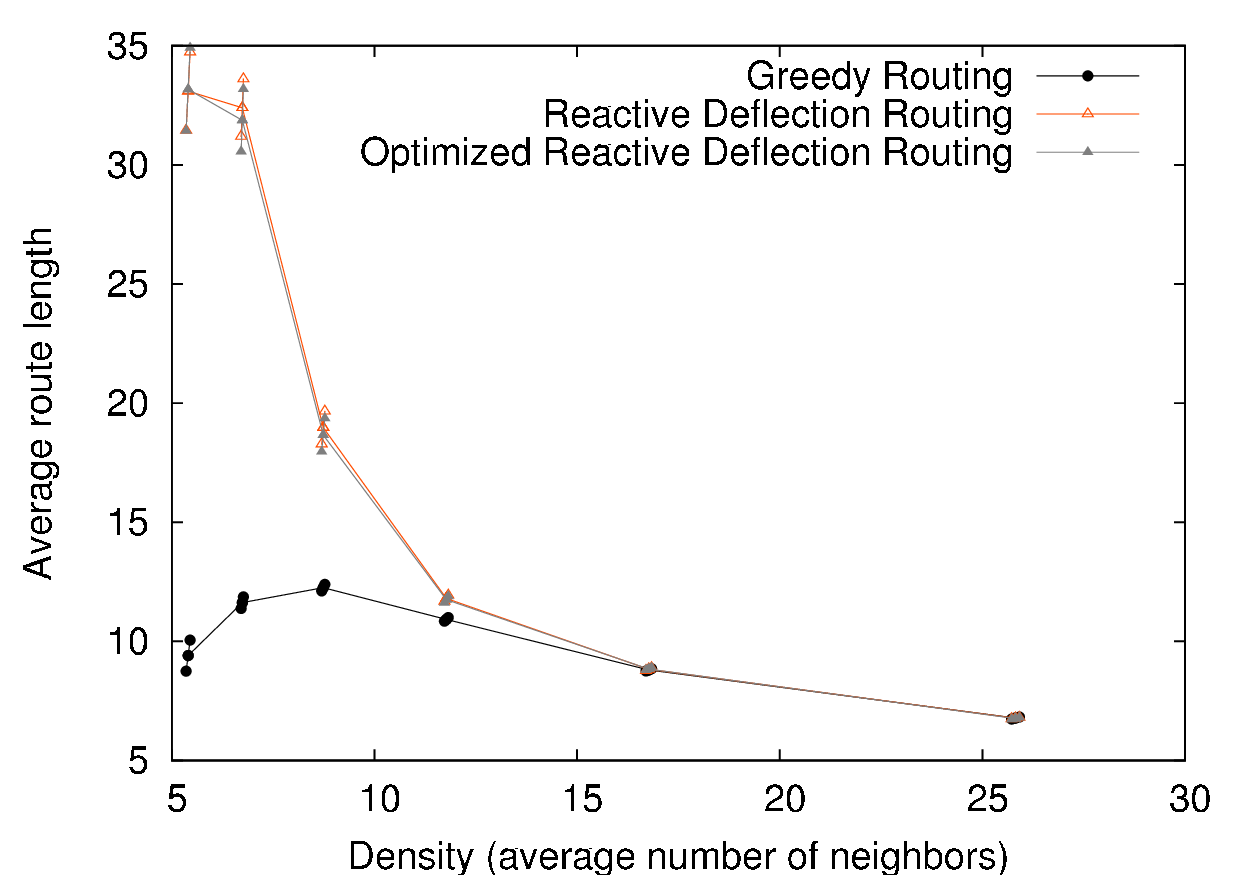}
		\label{fig:udg-route_length}
	}
	\subfigure[Stretch factor]{
		\includegraphics[width=9cm]{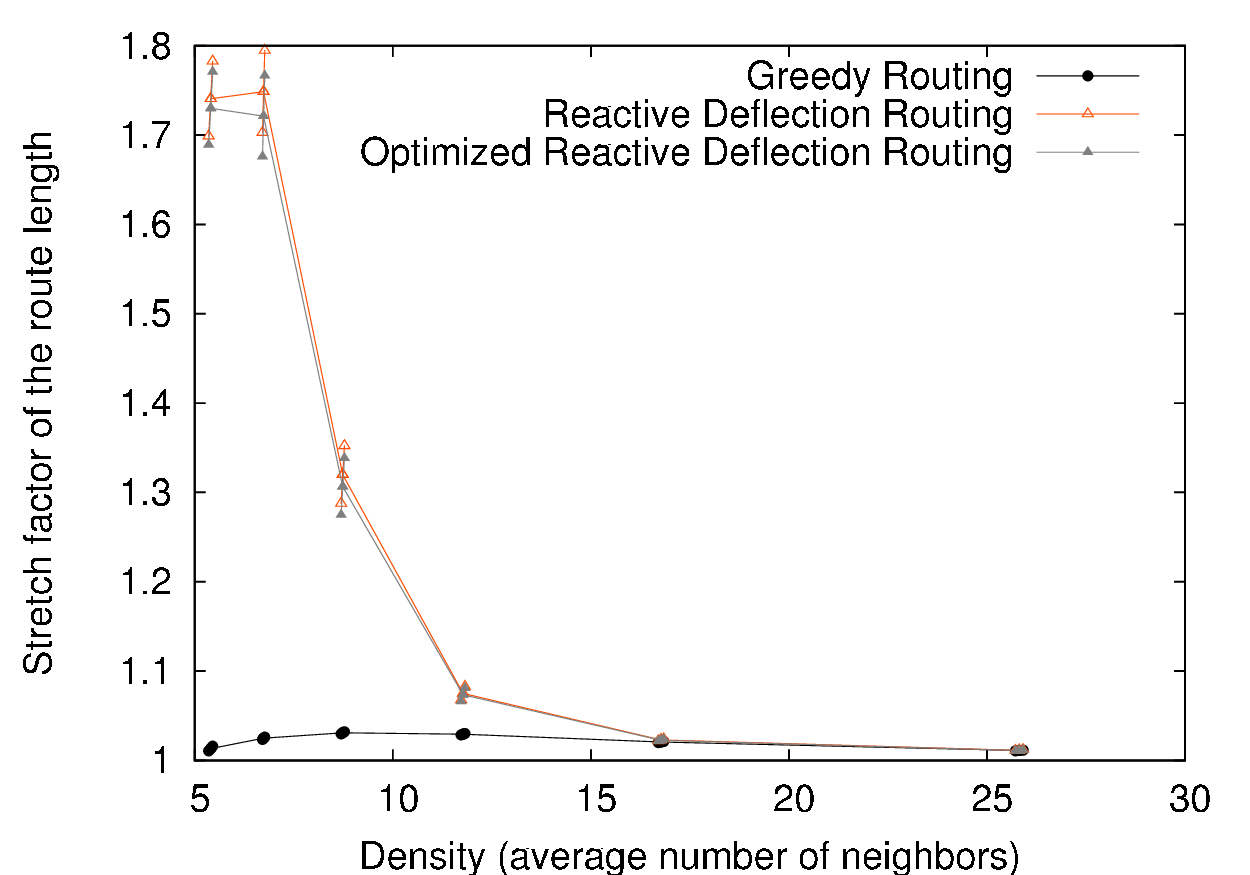}
		\label{fig:udg-stretch_factor}
	}
\caption{Route length under different routing schemes for UDG.}
\end{center}
\end{fig}

We compare our routing algorithm with greedy geographic routing to quantify the reduction of packet loss.  We use the classical version of greedy routing (the neighbor closest to the destination is chosen as next hop) since other versions do not show a significative improvement (smallest angle deviation, closest neighbor that is closer to the destination than myself). We mainly measure packet loss (the proportion of packets sent by a source that never reach the destination), route length (the average route length in hops for the delivered packets), and stretch factor (the average ratio of the route length for a packet and the length of the shortest route for the associated source/destination pair). We evaluated mainly the impact of density (average number of neighbors per node) on the routing performances. We plot the average values and the associated 95\% confidence intervals.

\subsection{Performance for Unit Disk Graphs}

In the first experiment, we have measured the route length obtained for deflection routing with different values of the k-neighborhood with density of 8 neighbors per node in Unit Disk Graphs (cf. Table~\ref{tab:kvalues}). We can remark that we quickly obtain a shorter route length with $k=3$. For larger values, deflection routing tends to overestimate the size of voids and increase the route length.

Then, the following experiment shows (cf. Figure~\ref{fig:udg-packet_loss}) that packet loss for greedy routing decreases with the increase of density: probability of having a large area without any node decreases so that voids are less probable to appear. However, more than 70\% of packets are lost for low density. On the contrary, the proposed routing scheme lowers packet loss: almost no packet is lost (less than 4\%) when density exceeds a small threshold (8 neighbors per node). Thus, nodes reroute less packets by means of reactive discovery so that the overhead is lower and delay improved. We can also notice that route length optimization has no impact on delivery ratio.

\begin{fig}
\begin{center}
	\subfigure[Route length.]{
		\includegraphics[width=9cm]{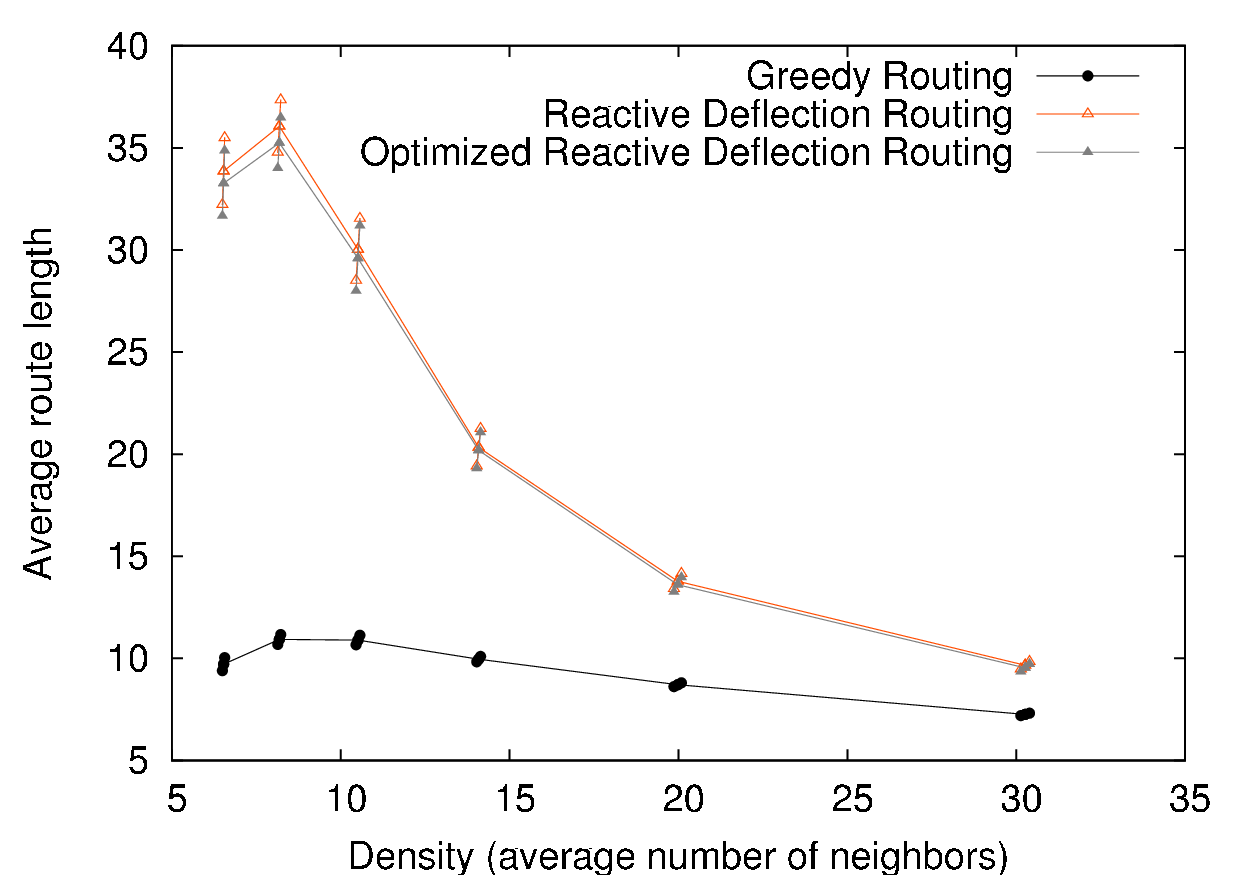}
		\label{fig:void_udg-route_length}
	}
	\subfigure[Packet losses.]{
		\includegraphics[width=9cm]{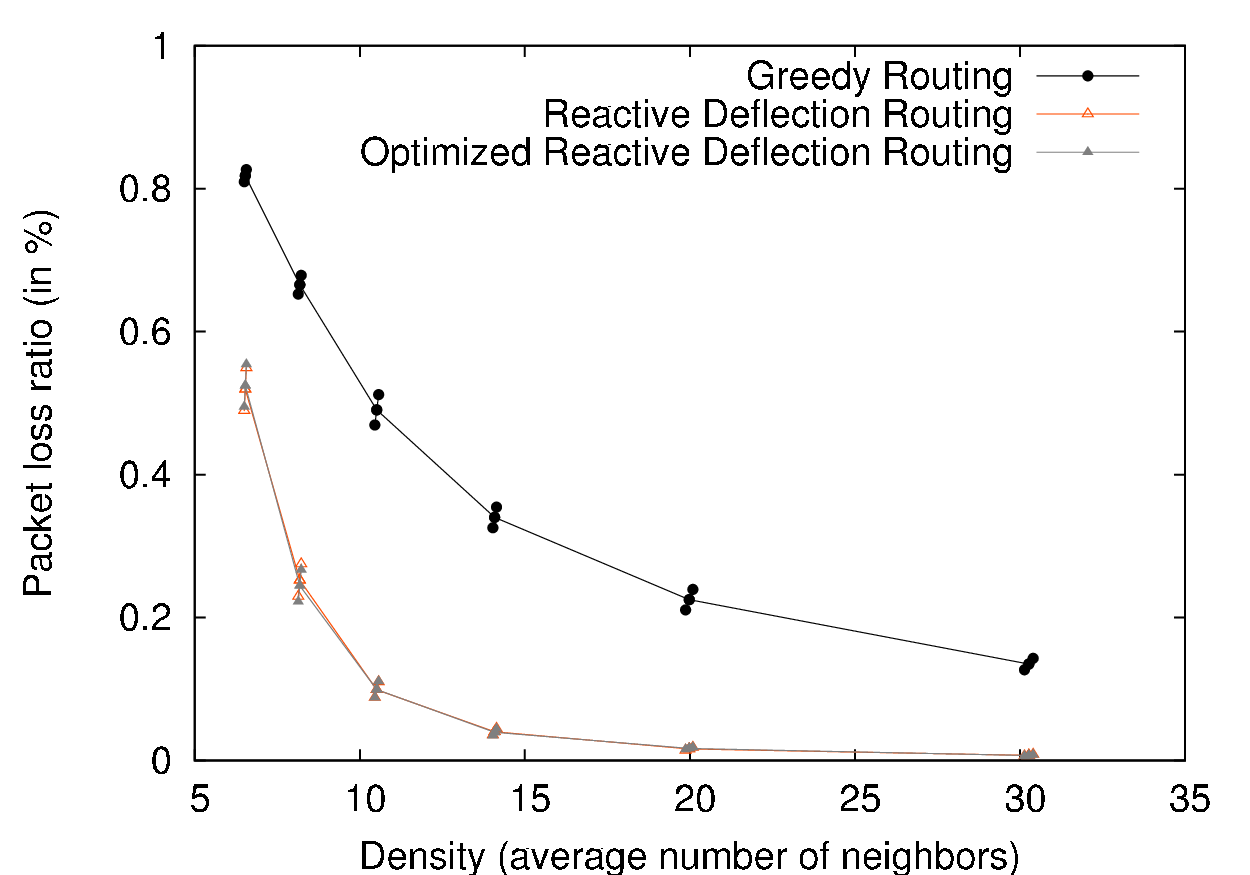}
		\label{fig:void_udg-packet_loss}
	}
	\caption{Performances of different routing schemes for a \proxgraph with one central void.}
	\label{fig:void_udg}
\end{center}
\end{fig}

We have also measured the route length (cf. Figure~\ref{fig:udg-route_length}). Greedy routing does not achieve to find routes when voids exist. Thus, the packet drop probability for greedy routing is larger when the destination is farther. Since the route length is only measured for delivered packets, this poor delivery ratio creates mechanically a lower average route length. To characterize its increase, we have measured the stretch factor (cf. Figure~\ref{fig:udg-stretch_factor}). We can observe that our optimized algorithm succeeds in slightly reducing the route
length. More importantly, route length optimization results in deviating packets from voids and decreasing the load on the void's borders. The stretch factor  is larger for low density since more voids are present and packets must be deflected and backtracked more often. The reader can note that greedy routing forwards one packet until it reaches a blocked node: this increases the load on these block nodes, even if they implement a void bypassing. Finally, the route length is reduced only for very low densities with the optimized version of deflection routing since blocked sectors interpolating is working only when voids are sufficiently large. Moreover, we could over-estimate the forbidden sector by adding a guarding angle around the blocked nodes: we would reduce the route length, but increase packet drops since we would over-estimate the presence of voids.

\subsection{Performance for a \proxgraph with one rectangular void}

We have evaluated packet loss rate in a \proxgraph with one central void (cf. Figure~\ref{fig:void_udg-packet_loss}). We can see that packet loss increases compared to unit disk graphs, particularly for high density: dead-ends are more probable. Moreover, since the graph is not UDG, a node may choose a next hop in a greedy way although it does not have any neighbor in the direction of the destination. Thus, to increase density it is not sufficient to surround voids. We can remark that with our algorithms we significantly reduce packet loss ratio.

Finally, we have measured the route length (cf. Figure~\ref{fig:void_udg-route_length}). We can remark the same trend as for UDG and for a \proxgraph with a void. Obviously, the route length is longer, because packets have to surround the rectangular void.

\subsection{Performance for more realistic channel conditions}

We have implemented greedy and deflection routing in ns (version 2.33) to test a non-ideal MAC and PHY layer. We have only consider 200 nodes, because of scalability limits of ns2. As above, we have placed one rectangular void in the center and all the nodes are placed randomly in the remaining simulation area. We have discarded all disconnected nodes. Finally, we have sequentially activated flows between random pairs of source and destination nodes. A flow sends 10 packets of 512 bytes with an inter-packet interval of 0.25s. In this way, we measure the ability of the routing protocol to discover a route rather than its robustness to the network load.

\begin{fig}
\begin{center}
	\subfigure[Total packet losses.]{
		\includegraphics[width=9cm]{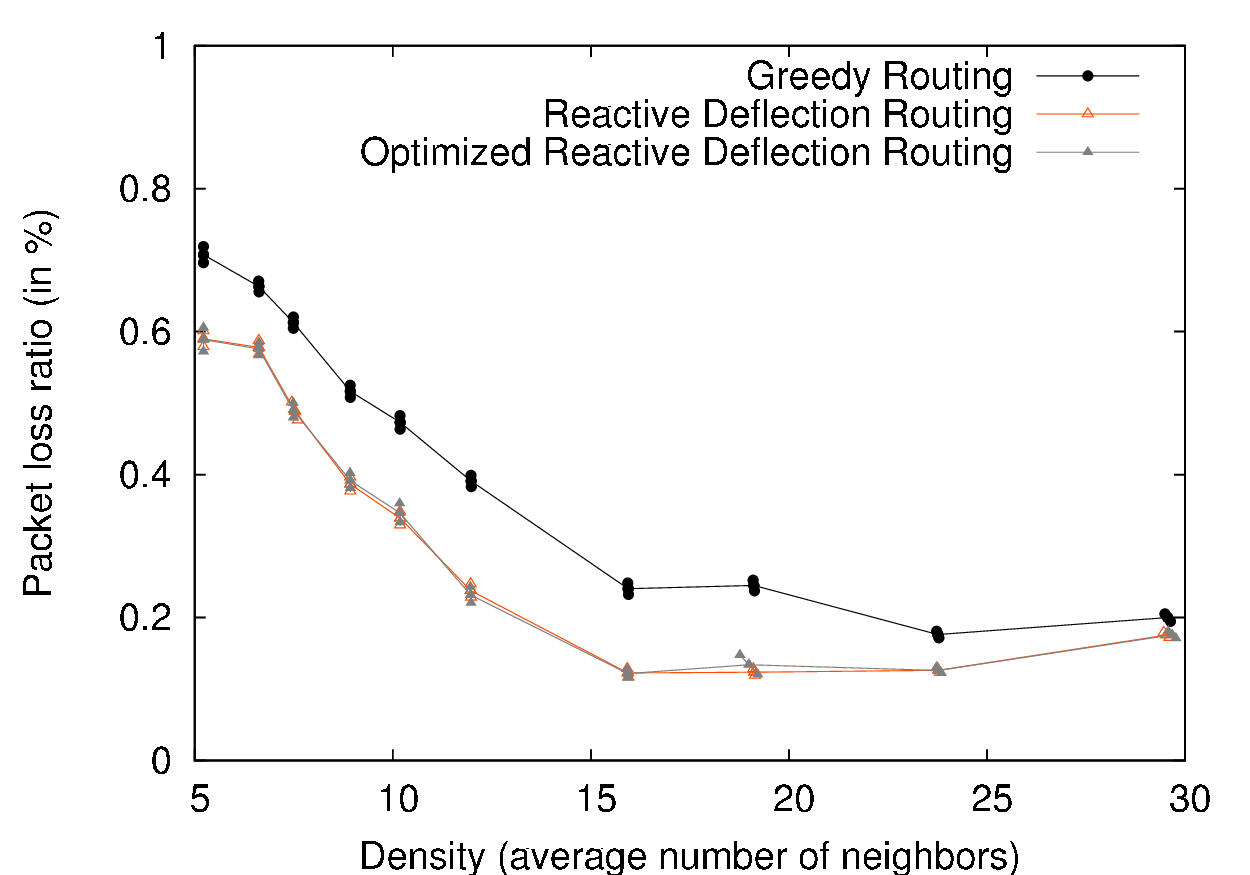}
		\label{fig: ns2-packet_loss_global}
	}
	\subfigure[Packet losses due to the absence of the next hop.]{
		\includegraphics[width=9cm]{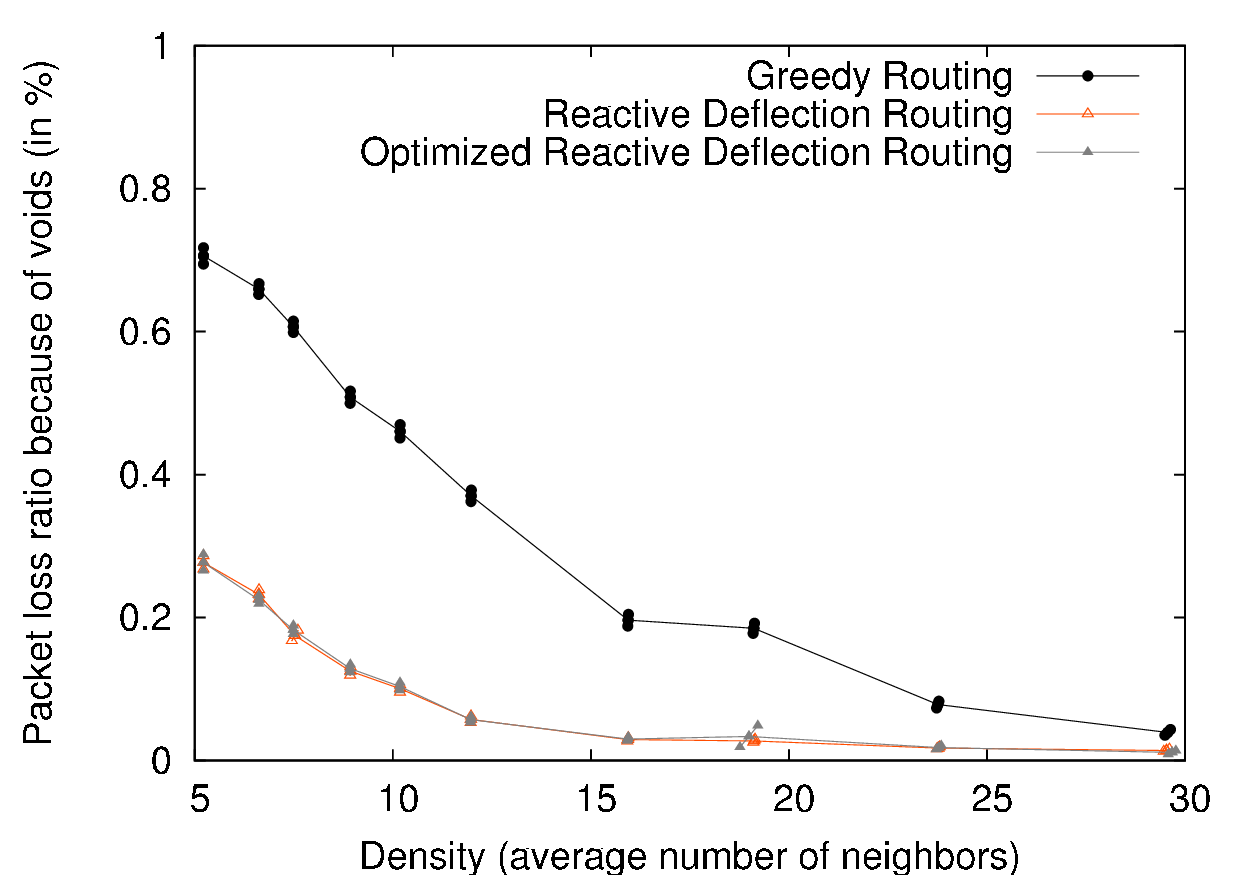}
		\label{fig: ns2-packet_loss_voids}
	}
	\vspace{-0.2cm}
	\caption{Packet loss ratio under different routing schemes in ns2.}
	\label{fig: ns2-packet_loss}
\end{center}
\end{fig}

We first report on the packet loss ratio for greedy and deflection routing (cf. Figure~\ref{fig: ns2-packet_loss}). We can remark that deflection routing achieves a lower loss rate than greedy routing: it discovers more routes. However, the MAC layer is now not ideal: packets can be dropped because of collisions or transmission errors, especially if the route is long. This explains the larger packet loss compared to the previous simulations. This effect also suggests that \ieee needs improvement for wireless mesh networks \cite{Iyer}.

Figure~\ref{fig: ns2-packet_loss_voids} represents packet losses only due to routing voids (i.e. there is no next hop according to the routing algorithm) that characterize the routing protocol and not the influence of the MAC layer (collisions, errors etc.). We can observe the same trends as for the \proxgraph (cf. Figure~\ref{fig:void_udg-packet_loss}): greedy routing suffers much more from voids than deflection routing. Finally, we also measure the route length (cf. Figure~\ref{fig: ns2-route_length}): although the route can be longer than for the ideal MAC and PHY layers, because for instance a
node could not discover a neighbor, deflection routing discovers routes that are not much longer than those in greedy routing. Besides, the optimized version of deflection routing becomes efficient in surrounding voids and reducing the route length in very sparse networks.

\begin{fig}
\begin{center}
	\includegraphics[width=9cm]{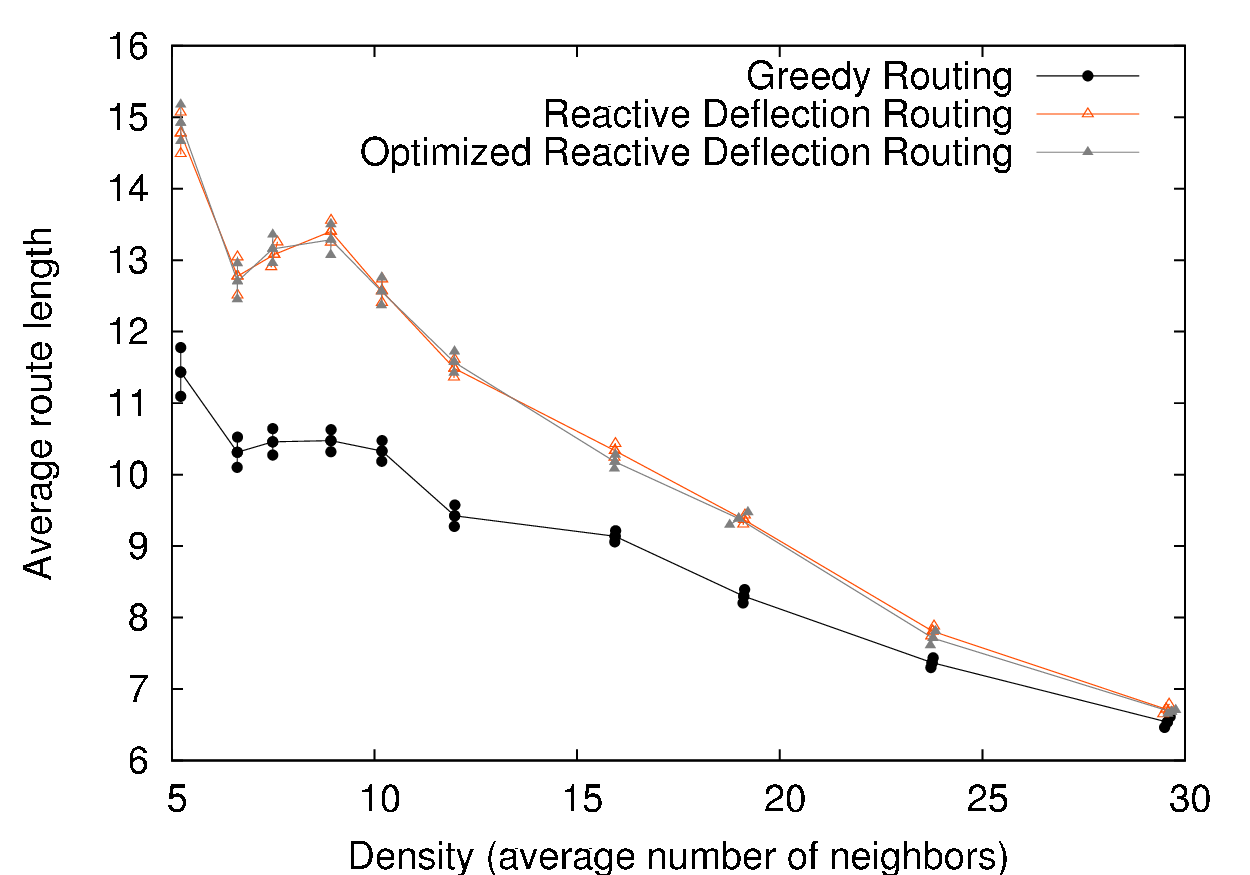}
	\vspace{-0.2cm}
	\caption{Route length under different routing schemes in ns2.}
	\label{fig: ns2-route_length}
\end{center}
\end{fig}


\section{Conclusion}

We have proposed a scheme for greedy geographical routing with reactive defect detection. The idea is to reactively detect blocked nodes and propagate the defect information by computing a set of blocked sectors. To reduce the route length and accelerate void detection in dense mesh networks, we have also proposed a method to extrapolate void location. Simulation results show good performance of the proposed methods: packet loss as well as the route length decrease compared to greedy routing.

\section*{Acknowledgments}
This work was partly supported by the European Commission project WIP under contract 2740, the French Ministry of Research project AIRNET under contract ANR-05-RNRT-012-01 and ARESA under contract ANR-05-RNRT-01703.

\bibliographystyle{abbrv}
\bibliography{georouting}

\end{document}